\documentclass[aps,prd,12pt,amssymb]{revtex4}

\begin{document}

\baselineskip=0.60cm

\newcommand{\ini}{\begin{equation}}
\newcommand{\fin}{\end{equation}}
\newcommand{\inir}{\begin{eqnarray}}
\newcommand{\finr}{\end{eqnarray}}
\newcommand{\inif}{\begin{figure}}
\newcommand{\finf}{\end{figure}}
\newcommand{\bc}{\begin{center}}
\newcommand{\ec}{\end{center}}

\def\ol{\overline}
\def\pa{\partial}
\def\ra{\rightarrow}
\def\ts{\times}
\def\df{\dotfill}
\def\bs{\backslash}
\def\dg{\dagger}

$~$

\hfill DSF-12/2003

\vspace{1 cm}

\title{BOUNDS FOR THE MASS OF THE HEAVIEST RIGHT-HANDED NEUTRINO IN SO(10) THEORIES}

\author{F. Buccella}
\author{D. Falcone}

\affiliation{Dipartimento di Scienze Fisiche,
Universit\`a di Napoli, Via Cintia, Napoli, Italy}

\begin{abstract}
\vspace{1cm}
\noindent
By relating the Dirac neutrino mass matrix to the mass of the charged
fermions and assuming that the product of the masses of the two lightest
left-handed neutrinos is of the order of $\Delta m^2_{sol}$, we derive,
within a leptogenesis scenario, a range of values for the mass of the
heaviest right-handed neutrino, centered around the scale of $B-L$
symmetry breaking in the $SO(10)$ theory with Pati-Salam intermediate
symmetry.
\end{abstract}

\maketitle

\newpage

Recently the neutrino oscillation \cite{p} solution to the solar neutrino
problem has been strongly supported by two experimental results:

a) the rate for the neutral current reaction $\nu + d \ra \nu + p + n$
measured at SNO \cite{sno}, where the number of active neutrinos is in
very good agreement with the flux of $\nu_e$ coming from the sun predicted
by the standard solar model \cite{ssm};

b) the disappearence of $\ol{\nu}_e$
found by KamLAND \cite{kamland}, consistent only with the MSW \cite{msw}
large mixing solution, which is also the preferred by the combined analysis
of the previous experiments \cite{prev}.

Together with the evidence for the oscillation of atmospheric neutrinos
\cite{atm} and with the limit by CHOOZ \cite{chooz} for the disappearence
of $\ol{\nu}_e$, it brings to a consistent description of the PMNS lepton
mixing matrix \cite{p,pmns}
\ini
U \simeq \left( \begin{array}{ccc}
\frac{\sqrt{3}}{2} & \frac{1}{2} & 0 \\
-\frac{1}{2 \sqrt2} & \frac{\sqrt3}{2 \sqrt2} & \frac{1}{\sqrt2} \\
\frac{1}{2 \sqrt2} & -\frac{\sqrt3}{2 \sqrt2} & \frac{1}{\sqrt2}
\end{array} \right) 
\fin
and the square mass differences \cite{fogli}
$$
\Delta m^2_{atm} \simeq 2.6 \cdot 10^{-3} ~\text{eV}^2,
$$
\ini
\Delta m^2_{sol} \simeq 7 \cdot 10^{-5} ~\text{eV}^2.
\fin
More severe limitations on the sum of neutrino masses are coming from
astrophysical observations \cite{wmap}
\ini
\sum |m_i| < 0.7 ~\text{eV}.
\fin

The smallness of neutrino masses with respect to charged fermions may be
understood in the framework of the seesaw mechanism \cite{ss}, which is
a strong indication in favour of $SO(10)$ unification \cite{soio},
where the existence of $\nu_R$ with very large Majorana masses is expected.
These particles are also an essential ingredient of the baryogenesis from
leptogenesis scenario \cite{fy}, for which a lower limit of about
$5 \cdot 10^8$ GeV for the mass of the lightest $\nu_R$ has been found
\cite{di}. Indeed, even with a perfect knowledge of the PMNS matrix
and of the square mass differences, there are several possibilities for the
right handed neutrino mass matrix $M_R$, since the Dirac mass matrix $M_D$
is unknown, as well as the mass of the lightest $\nu_L$ and the relative
phases of $m_i$.

In $SO(10)$, where the left-handed fermions of each family are classified in a
single representation (the spinorial {\bf 16}), we may try to relate
$M_D$ to the mass matrices of charged fermions. In the past years we have
stressed \cite{ab} that in a class of $SO(10)$-inspired models where one assumes
$M_{R33} \simeq 0$, the largest matrix element of $M_R$, namely $M_{R23}$,
takes a value of the order of the scale of $B-L$ symmetry breaking found in
the unified $SO(10)$ model \cite{b-l}. Here we want to consider the limits on the value of the mass
of the heaviest $\nu_R$ which follow from the hypothesis that, at the scale 
$M_Z$,
\ini
\text{Det} M_D= \frac{\text{Det} M_e \cdot \text{Det} M_u}{\text{Det} M_d},
\fin
where $M_e$, $M_u$ and $M_d$ are the mass matrices for charged leptons and
quarks. Eqn.(4) is valid with the Higgs doublet belonging to
{\bf 10} representations, but keeps its reliability for 
matrices like \cite{mm}
$$
M_u \simeq \left( \begin{array}{ccc}
0 & \sqrt{m_u m_c} & 0 \\
\sqrt{m_u m_c} & m_c & \sqrt{m_u m_t} \\
0 & \sqrt{m_u m_t} & m_t
\end{array} \right),
~~
M_d \simeq \left( \begin{array}{ccc}
0 & \sqrt{m_d m_s} & 0 \\
\sqrt{m_d m_s} & m_s & \sqrt{m_d m_b} \\
0 & \sqrt{m_d m_b} & m_b
\end{array} \right),
$$
\ini
M_D \simeq \frac{m_{\tau}}{m_b} \left( \begin{array}{ccc}
0 & \sqrt{m_u m_c} & 0 \\
\sqrt{m_u m_c} & -3 m_c & \sqrt{m_u m_t} \\
0 & \sqrt{m_u m_t} & m_t
\end{array} \right),
~
M_e \simeq \frac{m_{\tau}}{m_b} \left( \begin{array}{ccc}
0 & \sqrt{m_d m_s} & 0 \\
\sqrt{m_d m_s} & -3 m_s & \sqrt{m_d m_b} \\
0 & \sqrt{m_d m_b} & m_b
\end{array} \right),
\fin
where the factor $m_{\tau}/m_b$ comes from the renormalization from the
scale of quark-lepton unification to the electroweak scale and the factor
$-3$ in the 2-2 position, typical of Higgs transforming as a
$({\bf 15},{\bf 2},{\bf 2})$ of $SU(4) \times SU(2) \times SU(2)$, realizes
the Georgi-Jarlskog mechanism \cite{gj}. For simplicity we omit the CP
violating phase. We assume the normal hierarchy $|m_3| \gg |m_2|,|m_1|$.
From Eqn.(4) and the inverse seesaw formula
\ini
M_R \simeq -M_D^T M_L^{-1} M_D,
\fin
where $M_L$ is the effective neutrino mass matrix, we get
\ini
|\text{Det} M_R|=\left( 
\frac{m_e m_{\mu} m_{\tau} m_u m_c m_t}{m_d m_s m_b}
\right)^2
\frac{1}{|m_1 m_2 m_3|}=
1.6 \cdot 10^{30} k ~\text{GeV}^3,
\fin
where $k$ is the ratio $\Delta m^2_{sol}/|m_1 m_2|$ and we have considered
for the quark masses the values at the $M_Z$ scale \cite{fk}, since the
ratio $\text{Det} M_u/\text{Det} M_d$ is not modified by renormalization.
It should be stressed that Eqn.(4) implies a larger value for 
$\text{Det} M_D$ than for $\text{Det} M_e$. Moreover, Eqn.(7) implies for
the lightest and heaviest right-handed neutrino mass eigenvalues the inequalities
\ini
M_1 < k^{1/3}~1.2 \cdot 10^{10} < M_3.
\fin
The first part is consistent with the lower limit found for $M_1$ in the
baryogenesis from leptogenesis scenario \cite{di},
\ini
M_1 > 5 \cdot 10^8 ~\text{GeV},
\fin
since we have
\ini
k > \frac{7 \cdot 10^{-5}}{(2.3)^2 \cdot 10^{-2}}=1.3 \cdot 10^{-3},
\fin
where the upper limit for $|m_1 m_2|$ comes from Eqn.(3).
Then, from Eqns.(8)-(10) we get
\ini
1.1 \cdot 10^9 ~\text{GeV} < k^{1/3}~1.2 \cdot 10^{10} < M_3
< \frac{k~1.6 \cdot 10^{30}}{25 \cdot 10^{16}} =k~6.4 \cdot 10^{12} ~\text{GeV}.
\fin
In principle $k$ could be very large, since $|m_1|$ might be very small, but the
fact that
\ini
\frac{\Delta m^2_{sol}}{\Delta m^2_{atm}} \simeq 2.7 \cdot 10^{-2},
\fin
and the large mixing angles in the PMNS matrix, are
in favour of not so different neutrino masses. By taking $k=1$ the geometrical
value for the lower and upper limit for $M_3$ is $2.8 \cdot 10^{11}$ GeV,
very near to the value $2.7 \cdot 10^{11}$ GeV found for the scale of $B-L$
symmetry breaking in the $SO(10)$ gauge theory with the intermediate
$SU(4) \times SU(2) \times SU(2)$ symmetry \cite{b-l}.
The value for the scale of quark-lepton unification, smaller than the mass
of the lepto-quarks mediating proton decay, improves the
prediction for $m_b/m_{\tau}$ \cite{fk}. Conversely, by taking that value
for $M_3$ and the inequality (9) for $M_{1,2}$, one finds
\ini
|m_1 m_2| < (3 \cdot 10 ^{-2})^2 ~\text{eV}^2
\fin
which implies
\ini
\sum |m_i| < 0.12 ~\text{eV}.
\fin

In conclusion, the lower limit on $M_1$ in the baryogenesis via leptogenesis
scenario and the assumption (7), which is rather reliable in the framework of
$SO(10)$, and $k=1$, which is also reliable, since $\sqrt{\Delta m^2_{sol}}$
is less than one order smaller than $\sqrt{\Delta m^2_{atm}}$, estabilish a
range for $M_3$ centered around the value of the $B-L$ breaking scale in the
$SO(10)$ model with Pati-Salam intermediate symmetry \cite{ps}.
Our conclusion depends also on the assumption (4), which is not necessary true,
and from the hypothesis $k \sim 1$, which is reasonable. With smaller $|m_1|$
the lower limit found, which goes as $|m_1|^{-1/3}$, has a slow dependance
on $|m_1|$, while the upper limit, going as $|m_1|^{-1}$, varies more
rapidly.

\newpage

Finally, the result we found, which supports the $SO(10)$ theory with
Pati-Salam intermediate symmetry, does not exclude SUSY extensions,
where lower values of $M_R$ than the scale of $B-L$ breaking,
which is higher in that case, may be obtained from non-renormalizable
couplings \cite{nr}
$h \phi_{\bf 16} \phi_{\bf 16} \ol{\psi}_R^c \psi_L/M_P+\text{h.c.}$,
dumped by the Planck scale.

\end{document}